Quantifying the contribution of material and junction resistances in nano-networks

Cian Gabbett,[1*] Adam G. Kelly,[1*] Emmet Coleman,[1] Luke Doolan,[1] Tian Carey,[1] Kevin Synnatschke,[1] Shixin Liu,[1] Anthony Dawson,[1] Domhnall O'Suilleabhain,[1] Jose Munuera,[1] Eoin Caffrey,[1] John B. Boland,[1] Zdeněk Sofer,[2] Goutam Ghosh,[3] Sachin Kinge,[4] Laurens D.A. Siebbeles,[3] Neelam Yadav,[5] Jagdish K. Vij,[5] Muhammad Awais Aslam,[6] Aleksandar Matkovic[6] and Jonathan N. Coleman[1]**

[1]*School of Physics, CRANN & AMBER Research Centres, Trinity College Dublin, Dublin 2, Ireland*

[2]*Department of Inorganic Chemistry, University of Chemistry and Technology Prague, Technická 5, Prague 6, 166 28, Czech Republic*

[3]*Chemical Engineering Department, Delft University of Technology, Van der Maasweg 9, NL-2629 HZ Delft, The Netherlands*

[4]*Materials Research & Development, Toyota Motor Europe, B1930 Zaventem, Belgium*

[5]*Department of Electronic & Electrical Engineering, Trinity College Dublin 2, Ireland*

[6]*Chair of Physics, Department Physics, Mechanics and Electrical Engineering, Montanuniversität Leoben, Franz Josef Strasse 18, 8700 Leoben, Austria*

*These authors contributed equally, **colemaj@tcd.ie

ABSTRACT: Networks of nanowires and nanosheets are important for many applications in printed electronics. However, the network conductivity and mobility are usually limited by the inter-particle junction resistance, a property that is challenging to minimise because it is difficult to measure. Here, we develop a simple model for conduction in networks of 1D or 2D nanomaterials, which allows us to extract junction and nanoparticle resistances from particle-size-dependent D.C. resistivity data of conducting and semiconducting materials. We find junction resistances in porous networks to scale with nanoparticle resistivity and vary from 5 $\Omega$ for silver nanosheets to 25 G$\Omega$ for WS$_2$ nanosheets. Moreover, our model allows junction and nanoparticle resistances to be extracted from A.C. impedance spectra of semiconducting networks. Impedance data links the high mobility (~7 cm$^2$/Vs) of aligned networks of electrochemically exfoliated MoS$_2$ nanosheets to low junction resistances of ~670 k$\Omega$. Temperature-dependent impedance measurements allow us to quantitatively differentiate intra-nanosheet phonon-limited band-like transport from inter-nanosheet hopping for the first time.

*Introduction*

Printed electronic (PE) devices are increasingly important due to their flexibility, scalability, and cost-effectiveness.[1] Driven by their combination of solution-processibility and reasonable electrical performance, 0D, 1D, and 2D nanoparticles have now been widely explored as PE materials.[1,2] Printed networks of carbon nanotubes have shown great promise for use in transistors,[3,4] LEDs[5] and photodetectors[6] while metallic nanowire networks have been heavily studied as transparent electrodes,[7,8] EMI shields[9] and heating elements.[10] More recently, solution-processed networks of 2D materials such as graphene and $MoS_2$ have been investigated for a broad range of applications in all areas of (opto)electronics and energy storage.[11-13]

The success of this approach relies on exploiting the exceptional intrinsic properties of the individual nanoparticles (e.g., high conductivity or carrier mobility) when they are assembled into large-area networks. Conductive nanoparticles have been printed into networks with conductivities reaching ~$10^5$ S m$^{-1}$ for graphene,[12] ~$10^6$ S m$^{-1}$ for MXenes,[14] >$10^6$ S m$^{-1}$ for silver nanowires (AgNWs),[7,15] and >$10^7$ Sm$^{-1}$ for silver nanosheets (AgNS).[16] While the properties of conducting networks now approach those of the constituent nanoparticles, semiconducting networks have not kept up. Semiconducting carbon nanotube networks display mobilities <100 cm$^2$ V$^{-1}$ s$^{-1}$ [ref. 18] compared to >$10^5$ cm$^2$ V$^{-1}$ s$^{-1}$ for an individual nanotube,[17] while networks of solution-processed $MoS_2$ achieve ~10 cm$^2$ V$^{-1}$ s$^{-1}$ [refs. 12,18-20] compared to >100 cm$^2$ V$^{-1}$ s$^{-1}$ for mechanically exfoliated nanosheets.[21-23] The reason for these discrepancies is simple: almost all networks are limited by the junctions between particles, with semiconducting networks particularly junction-limited.[12,24]

Realising high-performance printed devices requires minimizing the junction resistance ($R_J$) relative to the nanoparticle resistance ($R_{NP}$). Achieving $R_J \ll R_{NP}$ would yield networks with properties approaching individual nanoparticles. Strategies to achieve this include optimising nanoparticle dimensions and deposition techniques,[12] or chemical cross-linking.[25,26] However, due to our inability to easily measure either the junction or the nanoparticle resistance in situ, assessing the progress of various strategies towards achieving $R_J < R_{NP}$ is difficult.

Despite their importance, the literature contains very little quantitative data on junction resistances. Although conductive-AFM[27] or micro-electrodes[28] can yield local information on both nanoparticle and junction resistances, these methods are unsuitable for large-area printed networks. Alternatively, although models exist linking junction resistance to network conductivity, they are too complex for routine utilisation.[29,30] The resultant lack of basic information has hindered printed device development and forced a reliance on trial-and-error for device optimisation.

Here, we develop a simple model relating the conductivity and mobility of nanoparticle networks to controllable nanoparticle parameters and network properties including junction resistance. We show that this model accurately describes experimental data for various nanomaterials and allows the extraction of both nanoparticle and junction resistances. We combine this model with impedance spectroscopy measurements to develop a powerful technique for simultaneously measuring both nanosheet and junction resistances within networks of semiconducting nanosheets.

*Model Development*

We utilise a classical, circuit-based approach to derive an equation for the resistivity of networks of 1D or 2D nanoparticles (e.g. nanowires or nanosheets), $\rho_{Net}$, in terms of properties of individual nanoparticles as well as the junction resistance, $R_J$, and network porosity, $P_{Net}$. Assuming a carrier traverses a well-defined current path (Fig. 1a) consisting of a linear array of nanoparticles (Fig. 1b), it must cross an inter-particle junction every time it traverses a nanoparticle. Current paths can then be modelled as linear arrays of nanoparticle-junction pairs, each pair represented by two resistors (Fig. 1b), describing the average nanoparticle ($R_{NP}$) and junction ($R_J$) resistances.

By relating the number of resistor pairs in a path to the channel length and the average distance travelled within each nanoparticle, one can estimate typical voltage drops across individual nanoparticles and junctions upon application of a voltage. These voltage drops yield the average transit times across individual nanoparticles and junctions. Combining these equations with an expression for the total transit time through the channel, one can obtain an equation for the network mobility:

$$\mu_{Net} \approx \frac{\mu_{NP}}{\left[1 + \frac{R_J}{R_{NP}}\right]\left[1 + \frac{2}{n_{NP} l_{NP} A_{NP}}\right]} \qquad (1)$$

where $\mu_{NP}$, $n_{NP}$, $l_{NP}$ and $A_{NP}$ are nanoparticle mobility, carrier density, length and cross-sectional area. This equation shows clearly that $\mu_{Net}$ depends on $R_J / R_{NP}$, which should be small to maximise mobility. Then combining an expression for network resistivity[12] ( $\rho_{Net}^{-1} = (1-P_{Net})n_{NP}e\mu_{Net})$ with expressions relating $A_{NP}$ and $R_{NP}$ to nanoparticle geometry, we can generate equations for $\rho_{Net}$ specific to 1D nanowires/nanotubes and 2D nanosheets:

$$\rho_{Net} \approx \frac{1}{(1-P_{Net})}\left[\rho_{NW} + \frac{\pi D_{NW}^2 R_J}{2l_{NW}}\right]\left[1 + \frac{8}{n_{NW}l_{NW}\pi D_{NW}^2}\right] \quad \text{(1D, 2a)}$$

$$\rho_{Net} \approx \frac{[\rho_{NS} + 2t_{NS}R_J]}{(1-P_{Net})}\left[1 + \frac{2}{n_{NS}t_{NS}l_{NS}^2}\right] \quad \text{(2D, 2b)}$$

where $D_{NW}$ and $t_{NS}$ are the nanowire diameter and nanosheet thickness and the geometry-specific subscripts, *NS* and *NW,* refer to nanosheet and nanowire, respectively. For large values of $n_{NP}$ such as those found for graphene, AgNS, or AgNWs, the second square-bracketed terms can be ignored.

*Measuring the change in network resistivity with nanoparticle dimensions*

Equations (2a) and (2b) suggest a rich array of size-dependent behaviour that has not yet been observed. For example, the appearance of nanosheet size parameters ($t_{NS}$, $l_{NS}$) in both denominator and numerator predicts a non-monotonic size-dependence with either a positive or negative $d\rho_{Net}/dt_{NS}$. To search for such behavior and to test the validity of equations (2a) and (2b), we produced inks of 1D AgNWs and four types of 2D nanosheets; graphene, $WS_2$, and $WSe_2$ (synthesised by liquid-phase exfoliation (LPE)[31]) and commercial AgNS. Each material was size-selected into fractions (Fig. 2a, c, d) which were spray-coated to produce a set of networks for electrical testing. All networks were thick enough to be in the bulk-like regime.[32]

The measured size-dependent D.C. resistivity is shown in Figures 3a-e for all five materials. Because $n_{NW}$ is large for AgNWs, equation (2a) predicts that $\rho_{Net}$ scales linearly with $(l_{NW})^{-1}$, behaviour that is clearly seen in Figure 3a. Nanosheets produced by LPE[33] and the AgNS display a roughly constant aspect ratio, $k_{NS}$, allowing us to reduce the number of variables in equation (2b) by replacing $t_{NS}$ with $t_{NS} = l_{NS}/k_{NS}$. Ignoring the final term in equation (2b) for graphene and AgNS, we now find that $\rho_{Net}$ should scale linearly with $l_{NS}$, as seen experimentally in Figures 3b-c. However, for semiconducting materials, $n_{NS}$ is low meaning the final term in equation (2b) must be considered and which predicts a resistivity-minimum at a specific nanosheet size. Figures 3d-e show $\rho_{Net}$ for $WS_2$ and $WSe_2$ which initially falls with increasing $l_{NS}$ before reaching a minimum, behaviour which is consistent with our non-intuitive prediction. That such a minima exists is important as it suggests the existence of an optimal nanosheet size.

Equations (2a) and (2b) describe our data well, including the counter-intuitive trends, with fitting yielding values for $R_J$ and $R_{NP}$ as shown in the figure panels. Converting $R_{NP}$ to $\rho_{NP}$ yields $\rho_{NW}$=1.5±0.2×10$^{-8}$ Ω m (AgNWs) and $\rho_{NS}$=7.2±3.9×10$^{-8}$ Ω m (AgNSs), similar to bulk silver (1.6×10$^{-8}$ Ω m). For graphene, we find $\rho_{NS}$=1.7±0.5×10$^{-5}$ Ω m, consistent with in-plane graphite (~10$^{-5}$-10$^{-6}$ Ω m)[34]. The $\rho_{NS}$ values for WS$_2$ (1.9±0.3 Ω m) and WSe$_2$ (0.35±0.07 Ω m) were consistent with previously reported values (0.6 Ω m [ref.[35]] and 0.1 Ω m [ref. [36]], respectively). The values of $R_J$ ranged from 5.2±0.7 Ω for AgNS to 24±2.3 GΩ for WS$_2$ and are consistent with previous estimates of $R_J$ ~3 Ω for AgNS,[16] 185 Ω for AgNWs,[37] ~10$^3$-10$^5$ W for graphene,[12] and ~10$^9$ Ω for MoS$_2$.[38] We note that $R_J$ scales with $\rho_{NP}$ (Figure 3f) and that $R_J/R_{NP}$-values were all >1 indicating that these networks were predominately junction-limited.

*A direct measurement of $R_J$ and $R_{NP}$ using impedance spectroscopy*

Measuring $R_{NP}$ and $R_J$ as described above is time-consuming. We propose that A.C. impedance spectroscopy, a powerful tool for device and materials characterisation,[39,40] can leverage the inherent junction capacitance to extract information about nanosheet and junction resistances. While similar proposals have been made previously for grain/grain-boundary[41,42,43] as well as 2D[44] systems, such measurements probe all junctions in all current paths, in practice yielding $R_{NP}$ and $R_J$ in arbitrary units (but *not* absolute values), limiting useful analysis.

To extract absolute values of $R_{NS}$ and $R_J$ for nanosheet networks, the impedance spectra of the network ($Z_{Net}$) must be converted to spectra representing the average nanosheet-junction pair ($Z_{NS-J}$) within the network. These $Z_{NS-J}$ spectra can then be analysed based on microscopic considerations (see below).

Equation (2b) relates the D.C. resistivity of a nanosheet network, $\rho_{Net}$, to $(R_{NS} + R_J)$, the resistance of the average nanosheet-junction pair. We propose the same scaling exists between the complex resistivity of the network, $\rho^*_{Net}$, and $Z_{NS-J}$ (where $\rho^*_{Net} = Z_{Net}A/L$, and $A$ and $L$ are the network cross-sectional area and channel length). This yields an equation which converts the real and imaginary parts of $\rho^*_{Net}$ to those representing the nanosheet-junction pair, once $P_{Net}$, $t_{NS}$, $l_{NS}$ and $n_{NS}$ are known:

$$Z_{NS-J} = \rho^*_{Net} \frac{(1-P_{Net})}{2t_{NS}} \left[1 + \frac{2}{n_{NS}t_{NS}l_{NS}^2}\right]^{-1} \quad (3)$$

We demonstrate this impedance approach using liquid-deposited networks of electrochemically exfoliated (EE) MoS$_2$ nanosheets ($l_{NS} \approx 1$ µm, $t_{NS} \approx 3.3$ nm) with low porosity[45] and large-area junctions[18] (Fig. 4a). While this is an intensively studied system due to its relatively high mobility (>1 cm$^2$ V$^{-1}$ s$^{-1}$ for printed networks)[18-20,46], the actual $R_{NS}$ and $R_J$ values are completely unknown. We first measure the (peak) field-effect mobility of these networks in a transistor geometry (Fig. 4b), obtaining an average of $\mu_{Net}$=6.6±0.6 cm$^2$ V$^{-1}$ s$^{-1}$, consistent with previous measurements.[18]

We then measured the real and imaginary parts of the complex resistivity as a function of frequency as shown in Fig. 4c. The low-frequency plateau of Re($\rho^*_{Net}$) yields a D.C. resistivity of $\rho_{Net}$ = 0.024 Ω m. Combining this value with the measured mobility gives a carrier density for the network of 3.8×10$^{23}$ m$^{-3}$, similar to previous values for EE-MoS$_2$.[20,47]

With these values now known, equation (3) can be used to convert the network impedance, $Z_{Net}$, into the impedance of the average nanosheet-junction pair, $Z_{NS-J}$. Figure 4d shows the real component of $Z_{NS-J}$, with the inset showing the imaginary component. In the A.C. domain, the nanosheet-junction pair can be described as the nanosheet resistance ($R_{NS}$) in series with a parallel resistor ($R_J$) and capacitor ($C_J$) representing the junction (Fig. 4c, inset), an arrangement referred to as the Randles circuit. The $Z_{NS-J}$ spectrum can be fitted using equations appropriate to the Randles circuit to yield values of $R_{NS}$, $R_J$, and $C_J$. We account for the inevitable distribution of junction resistances by fitting the data using a modified equation for the Randles circuit[48]:

$$\text{Re}\, Z_{NS-J}(\omega) = R_{NS} + \frac{R_J\left[1 + (\omega R_J C_J)^n \cos(n\pi/2)\right]}{1 + 2(\omega R_J C_J)^n \cos(n\pi/2) + (\omega R_J C_J)^{2n}} \quad (4)$$

where $n$ is an ideality factor that decreases from 1 as the distribution of $R_J C_J$ values in the network broadens.[49]

We find excellent fits yielding values of $R_J$=(2.9±0.1) MΩ, $R_{NS}$=(0.67±0.07) MΩ, $C_J$=(8.4±0.4)×10$^{-15}$ F, and n≈0.985. Over five devices on the same substrate, $R_{NS}$ typically varies by <20% with $R_J$ and $C_J$ showing larger spreads (SD/mean<60%) due to morphology variations. Here $R_J$ is >1000× lower than in Figures 3d-e for LPE nanosheets, while $R_J/R_{NS}$=4.4±0.2, meaning it is much less junction-limited than the LPE WS$_2$ and WSe$_2$ networks above. We can further analyse the nanosheet resistance by converting it to nanosheet resistivity, giving $\rho_{NS}$=(4.4±0.5)×10$^{-3}$ Ω m. As the meso-porosity of the network is extremely low (~0.02),[45] we assume the network carrier density is the same as the nanosheet carrier density which then implies a nanosheet mobility of 37±4 cm$^2$ V$^{-1}$ s$^{-1}$, reasonable for solution-processed MoS$_2$.[20,50]

We can support this result using several direct measurements. First, we used time-resolved pump-probe terahertz spectroscopy to determine the room-temperature A.C. mobility of photogenerated charge carriers. The observed mobility at a frequency of 1 THz is 40±2 cm$^2$ V$^{-1}$ s$^{-1}$, consistent with the value implied by impedance. Second, we performed field-effect mobility measurements on individual MoS$_2$ nanosheets (see Fig. 4e) obtaining a zero-gate-bias value of 42±6 cm$^2$ V$^{-1}$ s$^{-1}$, again consistent with our results. Combining this value with our $\mu_{Net}$ value and reformulating equation (1) as $R_J / R_{NS} \approx \mu_{NS} / \mu_{Net} - 1$, (neglecting the final term as $n_{NS}$ is large) allows us to estimate $R_J/R_{NS}$=5.3±1.4, again within error of the impedance result.

We then used *in-operando* frequency-modulated Kelvin Probe Force Microscopy (KPFM) measurements to map out the spatial distribution of the electrostatic potential across a network (Figs. 4f-g).[51-53] Between the biased and grounded electrodes, we find a combination of gradual decreases in potential within the nanosheets and well-defined potential drops at the junctions. By summing the potential drops at the junctions along the channel length, we extract the overall fraction of potential drop within nanosheets which yields a mean value of $R_J/R_{NS}$=10±4 (Fig. 4h). Although microstructural variations in similarly deposited networks will cause differences in $R_J$, we find these data to be extremely consistent supporting the validity of the impedance method.

*Using the impedance method: temperature dependence*

Impedance spectroscopy allows $R_J$ and $R_{NS}$ to be measured simultaneously under various circumstances. We demonstrate this by performing impedance measurements on networks of electrochemically exfoliated MoS$_2$ at various temperatures (Fig. 5). The low-frequency limit of the Re($\rho^*_{Net}$) spectrum (Fig. 5a) yields the network resistivity ($\rho_{Net}$) which is plotted versus 1/T in Figure 5b. Previous measurements on electrochemically exfoliated MoS$_2$ networks have shown $\rho_{Net}$ to follow activated behaviour around room temperature ($\rho = \rho_0 \exp(E_a / kT)$, $\rho_0$ and $E_a$ are constants) but 3D variable-range hopping[54] at lower temperatures ( $\rho = \rho_0 \exp[(T_0/T)^{1/4}]$, $\rho_0$ and $T_0$ are constants).[55] As shown in Figure 5b and 5b (inset), our data is consistent with this behaviour (fit constants in panel). However, this standard analysis cannot distinguish the respective contributions from the nanosheets and junctions. To decouple these properties, we first convert the network impedance spectra to Re($Z_{NS-J}$) and Im($Z_{NS-J}$) spectra, finding the well-defined temperature dependences shown in Figures 5c-d.

Fitting the Re($Z_{NS-J}$) spectrum to equation (4) yields values of $R_{NS}$, $R_J$, and $C_J$ for all temperatures as shown in Figures 5e-f. Figure 5e shows opposing temperature dependences for $R_J$ and $R_{NS}$ indicating hopping and band-like transport, respectively, with $R_{NS}/R_J$ increasing with temperature. Figure 5f shows a relatively small change in the junction capacitance, $C_J$, over the temperature range meaning the changes in the Re($Z_{NS-J}$) spectrum are dominated by changes in $R_J$. We find typical $C_J$ values of 6-8 fF, considerably smaller than a quantum capacitance (typically ~pF/mm$^2$)[56] but consistent with a geometric capacitance where $C_J = \varepsilon_r \varepsilon_0 A_J / l_J$. Taking $\varepsilon_r = 1$ and an inter-sheet distance of $l_J = 0.6$ nm [Ref. [20]], gives a junction area of $A_J = $ 0.4-0.5 μm$^2$, in excellent agreement with SEM measurements which yield $\langle A_J \rangle = 0.4$ μm$^2$ (Fig. 5f, inset).

To separately assess the transport mechanisms associated with the nanosheet and the junction, we examine the temperature dependence of $\rho_{NS}$, and $R_J$ in Figures 5g-h. Figure 5g shows $\rho_{NS}$ to scale as a power law, ($\rho_{NS} \propto T^\alpha$) with α ≈ 1.1, consistent with measurements on individual MoS$_2$ nanosheets (typically α=0.5-1.9)[57-59]. This behavior implies band-like transport, limited by phonon scattering,[60] which is commonly seen for individual MoS$_2$ nanosheets with high carrier densities [refs. [59,61,62]], and is also in agreement with the THz spectroscopy data (triangles).

As these networks are junction-limited, the temperature dependence of $R_J$ in Figure 5h is similar to that of $\rho_{Net}$ showing the same the transition from VRH to activated behaviour. We propose this behaviour is consistent with Miller-Abrahams-type[54] hopping between nanosheets such that

$$R_J \approx R_{J0} \exp(2l_J/a)\exp(E_a/kT) \tag{5a}$$

where $a$ is the localisation length. We also derived an alternative version of the 3D-VRH equation considering inter-nanosheet hopping from the conduction band-edge of one nanosheet to the conduction band of another yielding:

$$R_J \approx R_{J0} \exp\left(\frac{2l_J}{a}\right) \exp\left(\left[\frac{T_0}{T}\right]^{1/4}\right) \qquad (5b)$$

Where the constant $T_0$ is here given by $T_0 \sim 76\pi\hbar^2 d_0 / ka^3 m$ with $d_0$ being the monolayer thickness. Fitting the data in Figure 5h yields $E_a = 55\pm2$ meV and $T_0 = (471\pm37)\times10^3$ K, values which are solely associated with the junction. Our $E_a$ value is smaller than other reported values (in the absence of gating[55]) which is consistent with our low $R_J$ (equation (5a)) and relatively high network carrier mobility.[63] Combining $T_0$ with $m = 0.7m_e$ and $d_0 = 0.6$ nm, we calculate $a=0.7$ nm, similar to published values for $MoS_2$ (0.2-3 nm).[55,64-66] The most probable hopping distance was ~ 2 nm, again consistent with inter-sheet hopping.

In summary, our simple model for conduction in nanoparticle networks is extremely powerful in allowing junction and particle resistance to be extracted from experimental data and demonstrates that the network properties such as conductivity can be tuned by adjusting the material dimensions. It represents an important new tool for analysis of printed networks of technological important nanomaterials.

ACKNOWLEDGEMENTS: We acknowledge ERC grant FUTUREPRINT, the Graphene Flagship and the Horizon Europe project 2D-PRINTABLE. We have also received support from the Science Foundation Ireland (SFI) funded centre AMBER (SFI/12/RC/2278_P2) and availed of the facilities of the SFI-funded AML and ARM labs. T.C. acknowledges funding from a Marie Skłodowska-Curie Individual Fellowship "MOVE" (grant number 101030735, project number 211395, and award number 16883). A.M. acknowledges support from the European Research Council Starting Grant POL_2D_PHYSICS (101075821) and the Austrian Science Fund Y1298-N START Prize. G.G., S.K. and L.D.A.S. received funding from the Netherlands Organization for Scientific Research (NWO) in the framework of the Materials for sustainability and from the Ministry of Economic Affairs in the framework of the PPP allowance. Z.S. was supported by ERC-CZ program (project LL2101) from Ministry of Education Youth and Sports (MEYS) and acknowledges laser infrastructure from project reg. No. CZ.02.1.01/0.0/0.0/15_003/0000444 financed by the EFRR.

Figures:

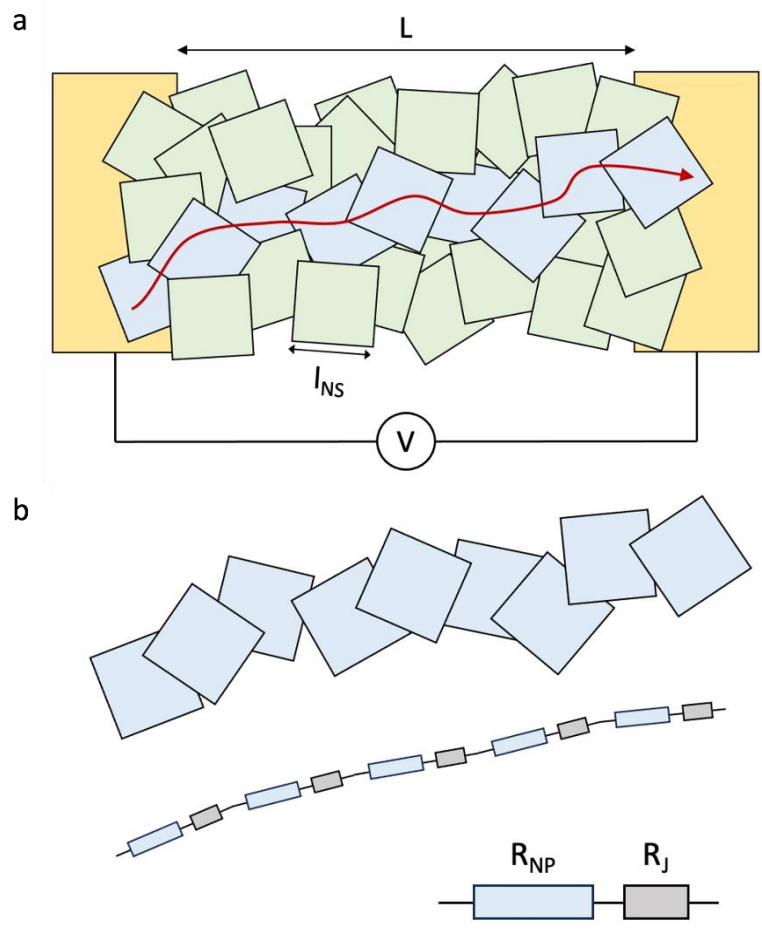

Figure 1: **Model Schematics**. **a**) Schematic illustrating a nanosheet network connected to two electrodes with channel length, $L$. The nanosheet lateral size is $l_{NS}$. While this schematic depicts a nanosheet network, a similar diagram could easily be produced to represent nanowire network. **b**) A single conducting path consisting of a linear array of nanosheets spanning the entire channel length. This conducting path can be considered as a chain of resistor pairs, with each pair consisting of a resistance representing a nanosheet and one representing the inter-sheet junction.

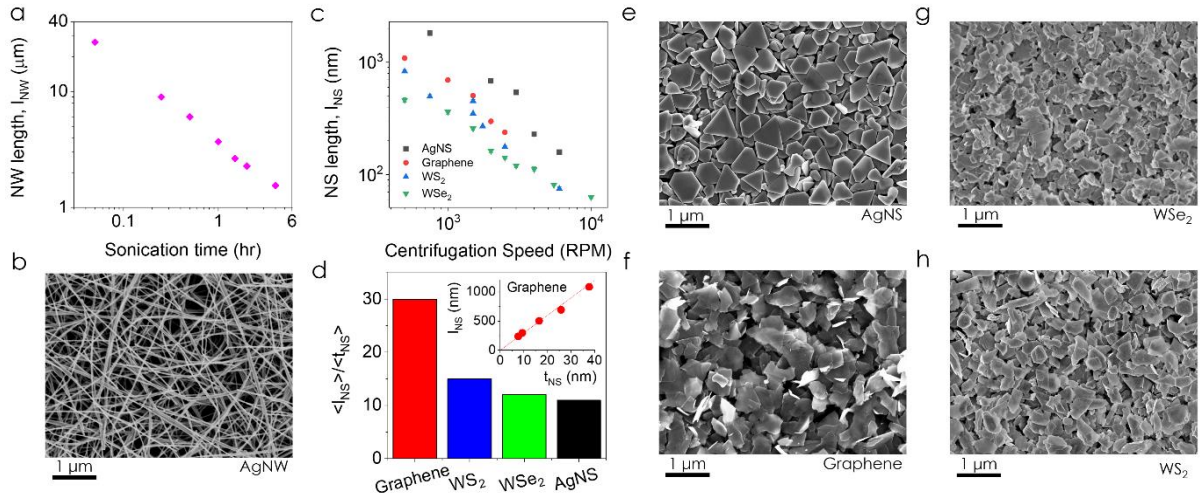

Figure 2: **Microscopic analysis of nanoparticles and networks**. **a**) Mean length of AgNWs size-selected by sonication-induced scission as a function of sonication time. **b**) SEM image of the top surface of a spray-cast network of AgNWs. **c**) Mean sizes of centrifuge-fractionated 2D materials: AgNS, graphene, $WS_2$, and $WSe_2$ nanosheets, plotted versus centrifugation speed. **d**) Mean aspect ratio of all four 2D materials, averaged over all fractions. Inset: Exemplary data showing linear scaling between nanosheet length and thickness over five size-selected fractions of LPE graphene, consistent with aspect ratio of 30. SEM images of the top surface of spray-cast networks of AgNS (e), graphene (f), $WSe_2$ (g), and $WS_2$ (h) nanosheets.

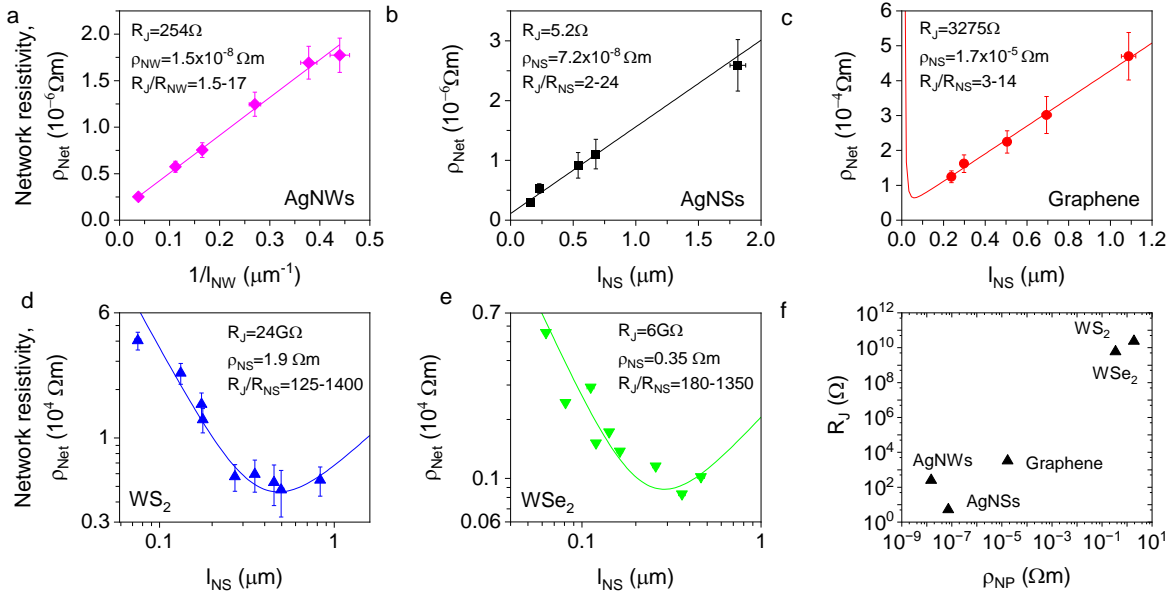

Figure 3: **Dependence of network resistivity on nanoparticle dimensions**. **a)** Resistivity of spray-cast silver nanowire networks versus inverse nanowire length. The line is a fit to equation (2a). **b-e)** Resistivity of spray-cast nanosheet networks versus nanosheet length for networks of: AgNS (b), graphene (c), WS$_2$ (d), and WSe$_2$ (e). In (b)-(e), the lines represent fits to equation (2b). In (a), (b), and (c) the carrier density is large allowing us to neglect the second square-bracketed term in the fitting equations. The behaviour in (b) and (c) is counter-intuitive as the general expectation is that smaller nanosheets lead to higher resistivity. Fitting yields values of $R_J$ and $\rho_{NP}$. The latter parameter, combined with nanoparticle dimensions yields $R_{NP}$. Values of $R_J$ and $\rho_{NP}$, as well as ranges of $R_J/R_{NP}$, are given in the panels. **f)** Junction resistance, $R_J$, plotted versus nanoparticle resistivity, $\rho_{NP}$, demonstrating scaling.

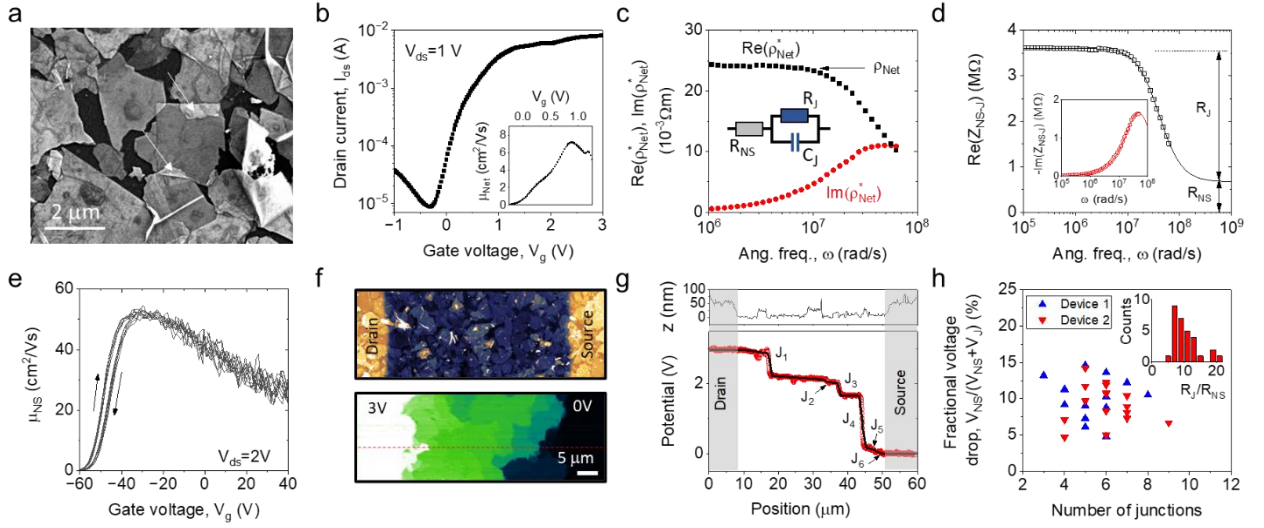

Figure 4: **Identification of nanosheet and junction resistances. a**) An SEM image of the surface of a network of electrochemically exfoliated (EE) MoS$_2$ nanosheets. The arrows point to two well-defined junctions. **b**) Field effect transfer curve for an electrolytically gated EE MoS$_2$ network very similar to that in (a). Inset: Gate-voltage-dependent mobility for an EE MoS$_2$ network nanosheet. Averaging over four such devices yields a mean (peak) mobility of $\mu_{Net}$=6.6±0.6 cm$^2$/Vs. **c**) Real and imaginary parts of the complex resistivity, both plotted as a function of angular frequency, ω, measured for a network of electrochemically exfoliated EE MoS$_2$ nanosheets similar to that in (a). Inset: The circuit element representing a nanosheet-junction pair. Here, $R_{NS}$ is the nanosheet resistance while $R_J$ and $C_J$ are the junction resistance and capacitance respectively. **d**) The real part of the impedance of a nanosheet-junction pair (Re($Z_{NS-J}$)) plotted versus ω. The data has been fitted using equation (4). The relationship between this curve and the junction and nanosheet resistances is shown by the arrows. Inset: -Im($Z_{NS-J}$) plotted as a function of ω. The solid line is a fit, see SI S11 for equation and fit parameters. **e**) Gate voltage dependent mobility for a representative individual EE MoS$_2$ nanosheet. **f**) Topographic AFM image (top) and *in-operando* KPFM image (bottom) of a section of a channel comprising an EE MoS$_2$ network. **g**) Topographic line profile (top) and potential profile (bottom) associated with the red dashed line in (f). In this section of channel, 6 sharp drops associated with inter-sheet junctions can be seen, labeled as J$_1$ to J$_6$. The flat regions represent the gradual drop of potential across nanosheets. The black line represents fits to the linear regions. **h**) Fraction of voltage dropped across nanosheets in a given section of channel plotted as a function of the number of junctions observed in that section. Inset: Histogram of $R_J/R_{NS}$ values as calculated from the fractional voltage drops in the main panel.

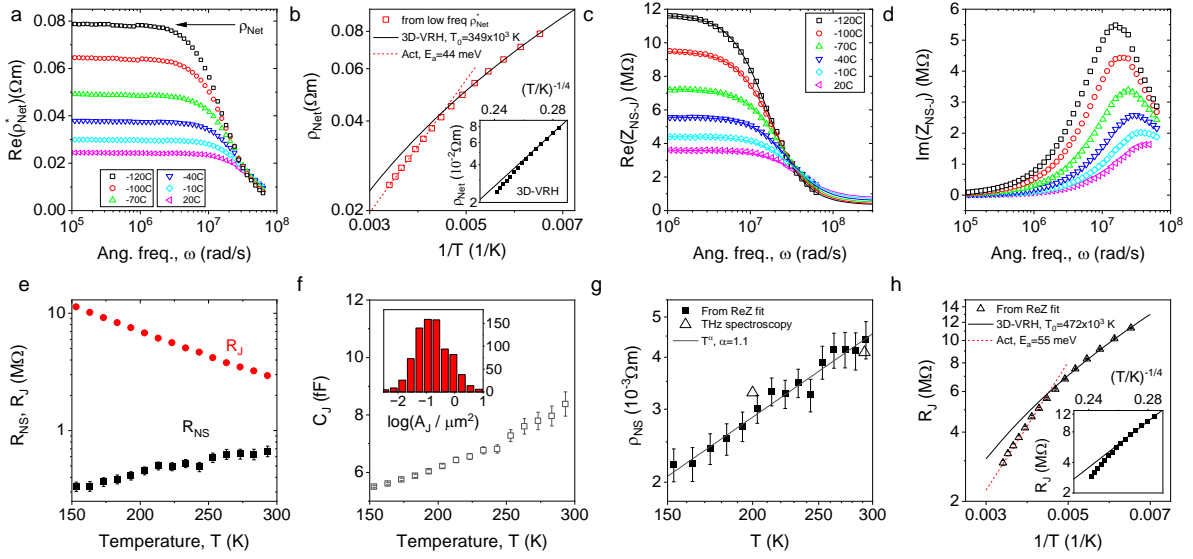

Figure 5: **Measurement of network, nanosheet, and junction transport using impedance spectroscopy**. **a**) Real part of the complex A.C. resistivity plotted as a function of angular frequency, ω, measured for a network of electrochemically exfoliated EE MoS$_2$ nanosheets at a range of temperatures. The arrow indicates that the D.C. network resistivity was found from $\rho_{Net} = \mathrm{Re}(\rho^*_{Net})_{\omega\rightarrow 0}$. **b**) D.C. network resistivity plotted as a function of 1/T and T$^{-1/4}$ (inset). The solid line is an activated fit while the dashed line is a fit to the 3D-VRH model. Fit parameters are given in the panel. **c-d**) Real (c) and imaginary (d) parts of the impedance spectrum of a single (average) nanosheet junction pair, measured at various temperatures. The curves in (c) are fitted using equation (4). **e**) $R_{NS}$ and $R_J$, extracted from fits to Re($Z_{NS-J}(\omega)$) spectra. **f**) Junction capacitance plotted versus temperature. Inset: Junction area histogram, measured from SEM images (plotted as log($A_J$/μm$^2$)). This distribution showed $\langle A_J \rangle$=0.41 μm$^2$. **g**) Resistivity of an (average) individual nanosheet, extracted from $R_{NS}$ ($\rho_{NS} \approx 2R_{NS}t_{NS}$), plotted as function of temperature. The solid line is a power law with exponent 1.1. The triangles represent the THz mobility of the nanosheets converted into resistivity using the measured carrier density of 3.8×10$^{23}$ m$^{-3}$. **h**) Junction resistance, $R_J$, plotted as a function of 1/T and T$^{-1/4}$ (inset). The solid line is an activated-behaviour fit while the dashed line is a fit to the 3D-VRH model. Fit parameters are given in the panel.

# References

1. Chandrasekaran, S., Jayakumar, A. & Velu, R. A Comprehensive Review on Printed Electronics: A Technology Drift towards a Sustainable Future. *Nanomaterials* **12**, (2022).
2. Gulzar, U., Glynn, C. & O'Dwyer, C. Additive manufacturing for energy storage: Methods, designs and material selection for customizable 3D printed batteries and supercapacitors. *Current Opinion in Electrochemistry* **20**, 46-53, (2020).
3. Schiessl, S. P. *et al.* Polymer-Sorted Semiconducting Carbon Nanotube Networks for High-Performance Ambipolar Field-Effect Transistors. *ACS Appl. Mater. Interfaces* **7**, 682-689, (2015).
4. Lu, S. H., Smith, B. N., Meikle, H., Therien, M. J. & Franklin, A. D. All-Carbon Thin-Film Transistors Using Water-Only Printing. *Nano Lett.* **23**, 2100-2106, (2023).
5. Graf, A., Murawski, C., Zakharko, Y., Zaumseil, J. & Gather, M. C. Infrared Organic Light-Emitting Diodes with Carbon Nanotube Emitters. *Adv. Mater.* **30**, 1706711, (2018).
6. Richter, M., Heumüller, T., Matt, G. J., Heiss, W. & Brabec, C. J. Carbon Photodetectors: The Versatility of Carbon Allotropes. *Adv. Energy Mater.* **7**, 1601574, (2017).
7. De, S. *et al.* Silver Nanowire Networks as Flexible, Transparent, Conducting Films: Extremely High DC to Optical Conductivity Ratios. *ACS Nano* **3**, 1767-1774, (2009).
8. Hu, L., Kim, H. S., Lee, J.-Y., Peumans, P. & Cui, Y. Scalable Coating and Properties of Transparent, Flexible, Silver Nanowire Electrodes. *ACS Nano* **4**, 2955-2963, (2010).
9. Lin, S. *et al.* Room-temperature production of silver-nanofiber film for large-area, transparent and flexible surface electromagnetic interference shielding. *npj Flexible Electronics* **3**, 6, (2019).
10. Zhu, Y. *et al.* Flexible Transparent Electrodes Based on Silver Nanowires: Material Synthesis, Fabrication, Performance, and Applications. *Advanced Materials Technologies* **4**, 1900413, (2019).
11. Witomska, S., Leydecker, T., Ciesielski, A. & Samori, P. Production and Patterning of Liquid Phase-Exfoliated 2D Sheets for Applications in Optoelectronics. *Adv. Funct. Mater.* **29**, (2019).
12. Kelly, A. G., O'Suilleabhain, D., Gabbett, C. & Coleman, J. N. The electrical conductivity of solution-processed nanosheet networks. *Nature Reviews Materials* **7**, 217-234, (2022).
13. Zhu, X. X. *et al.* Hexagonal Boron Nitride-Enhanced Optically Transparent Polymer Dielectric Inks for Printable Electronics. *Adv. Funct. Mater.* **30**, (2020).
14. He, P. *et al.* Screen-Printing of a Highly Conductive Graphene Ink for Flexible Printed Electronics. *ACS Appl. Mater. Interfaces* **11**, 32225-32234, (2019).
15. Sannicolo, T. *et al.* Metallic Nanowire-Based Transparent Electrodes for Next Generation Flexible Devices: a Review. *Small* **12**, 6052-6075, (2016).
16. Kelly, A. G. *et al.* Highly Conductive Networks of Silver Nanosheets. *Small* **18**, (2022).
17. Zhou, X. J., Park, J. Y., Huang, S. M., Liu, J. & McEuen, P. L. Band structure, phonon scattering, and the performance limit of single-walled carbon nanotube transistors. *Phys. Rev. Lett.* **95**, (2005).
18. Carey, T. *et al.* High-Mobility Flexible Transistors with Low- Temperature Solution-Processed Tungsten Dichalcogenides. *ACS Nano*, (2023).

19  Song, O. *et al.* All inkjet-printed electronics based on electrochemically exfoliated two-dimensional metal, semiconductor, and dielectric. *Npj 2d Materials and Applications* **6**, (2022).
20  Lin, Z. Y. *et al.* Solution-processable 2D semiconductors for high-performance large-area electronics. *Nature* **562**, 254-+, (2018).
21  Radisavljevic, B., Radenovic, A., Brivio, J., Giacometti, V. & Kis, A. Single-layer MoS2 transistors. *Nat. Nanotechnol.* **6**, 147-150, (2011).
22  Perera, M. M. *et al.* Improved Carrier Mobility in Few-Layer MoS2 Field-Effect Transistors with Ionic-Liquid Gating. *ACS Nano* **7**, 4449-4458, (2013).
23  Liu, H. & Ye, P. D. MoS2 Dual-Gate MOSFET With Atomic-Layer-Deposited Al2O3 as Top-Gate Dielectric. *IEEE Electron Device Lett.* **33**, 546-548, (2012).
24  Zorn, N. F. & Zaumseil, J. Charge transport in semiconducting carbon nanotube networks. *Applied Physics Reviews* **8**, (2021).
25  Ippolito, S. *et al.* Covalently interconnected transition metal dichalcogenide networks via defect engineering for high-performance electronic devices. *Nat. Nanotechnol.* **16**, 592-598, (2021).
26  Ippolito, S. *et al.* Unveiling Charge-Transport Mechanisms in Electronic Devices Based on Defect-Engineered MoS2 Covalent Networks. *Adv. Mater.*, (2023).
27  Nirmalraj, P. N., Lyons, P. E., De, S., Coleman, J. N. & Boland, J. J. Electrical connectivity in single-walled carbon nanotube networks. *Nano Lett.* **9**, 3890-3895, (2009).
28  Bellew, A. T., Manning, H. G., da Rocha, C. G., Ferreira, M. S. & Boland, J. J. Resistance of Single Ag Nanowire Junctions and Their Role in the Conductivity of Nanowire Networks. *ACS Nano* **9**, 11422-11429, (2015).
29  Forro, C., Demko, L., Weydert, S., Voros, J. & Tybrandt, K. Predictive Model for the Electrical Transport within Nanowire Networks. *ACS Nano* **12**, 11080-11087, (2018).
30  O'Callaghan, C., da Rocha, C. G., Manning, H. G., Boland, J. J. & Ferreira, M. S. Effective medium theory for the conductivity of disordered metallic nanowire networks. *Phys. Chem. Chem. Phys.* **18**, 27564-27571, (2016).
31  Bonaccorso, F., Bartolotta, A., Coleman, J. N. & Backes, C. 2D-Crystal-Based Functional Inks. *Adv. Mater.* **28**, 6136-6166, (2016).
32  De, S., King, P. J., Lyons, P. E., Khan, U. & Coleman, J. N. Size Effects and the Problem with Percolation in Nanostructured Transparent Conductors. *ACS Nano* **4**, 7064-7072, (2010).
33  Backes, C. *et al.* Equipartition of Energy Defines the Size-Thickness Relationship in Liquid-Exfoliated Nanosheets. *ACS Nano* **13**, 7050-7061, (2019).
34  Nakamura, S., Miyafuji, D., Fujii, T., Matsui, T. & Fukuyama, H. Low temperature transport properties of pyrolytic graphite sheet. *Cryogenics* **86**, 118-122, (2017).
35  Agarwal, M. K., Nagireddy, K. & Patel, P. D. Electrical-Properties of Tungstenite (WS2) Crystals. *Kristall Und Technik-Crystal Research and Technology* **15**, K65-K67, (1980).
36  Upadhyayula, L. C., Loferski, J. J., Wold, A., Giriat, W. & Kershaw, R. Semiconducting Properties of Single Crystals Of N And P-Type Tungsten Diselenide (WSe2). *J. Appl. Phys.* **39**, 4736-+, (1968).
37  Song, T. B. *et al.* Nanoscale Joule Heating and Electromigration Enhanced Ripening of Silver Nanowire Contacts. *ACS Nano* **8**, 2804-2811, (2014).
38  Biccai, S. *et al.* Negative Gauge Factor Piezoresistive Composites Based on Polymers Filled with MoS2 Nanosheets. *ACS Nano* **13**, 6845-6855, (2019).

39     Lazanas, A. C. & Prodromidis, M. I. Electrochemical Impedance Spectroscopy-A Tutorial. *Acs Measurement Science Au* **3**, 162-193, (2023).

40     Yim, C., McEvoy, N., Kim, H. Y., Rezvani, E. & Duesberg, G. S. Investigation of the Interfaces in Schottky Diodes Using Equivalent Circuit Models. *ACS Appl. Mater. Interfaces* **5**, 6951-6958, (2013).

41     Irvine, J. T. S., Sinclair, D. C. & West, A. R. Electroceramics: Characterization by Impedance Spectroscopy. *Adv. Mater.* **2**, 132-138, (1990).

42     Gerstl, M. *et al.* The separation of grain and grain boundary impedance in thin yttria stabilized zirconia (YSZ) layers. *Solid State Ionics* **185**, 32-41, (2011).

43     Fleig, J. & Maier, J. The impedance of ceramics with highly resistive grain boundaries: validity and limits of the brick layer model. *J. Eur. Ceram. Soc.* **19**, 693-696, (1999).

44     Moore, D. C. *et al.* Ultrasensitive Molecular Sensors Based on Real-Time Impedance Spectroscopy in Solution-Processed 2D Materials. *Adv. Funct. Mater.* **32**, (2022).

45     Cian Gabbett *et al.* 3D-imaging of Printed Nanostructured Networks using High-resolution FIB-SEM Nanotomography. *arXiv:2301.11046*, (2023).

46     Kim, J. *et al.* All-Solution-Processed Van der Waals Heterostructures for Wafer-Scale Electronics. *Adv. Mater.* **34**, (2022).

47     Gao, X. *et al.* High-mobility patternable MoS2 percolating nanofilms. *Nano Research* **14**, 2255-2263, (2021).

48     Heil, T. & Jossen, A. Continuous approximation of the ZARC element with passive components. *Meas. Sci. Technol.* **32**, (2021).

49     Boukamp, B. A. Distribution (function) of relaxation times, successor to complex nonlinear least squares analysis of electrochemical impedance spectroscopy? *Journal of Physics-Energy* **2**, (2020).

50     Kelly, A. G. *et al.* All-printed thin-film transistors from networks of liquid-exfoliated nanosheets. *Science* **356**, 69-72, (2017).

51     Palermo, V., Palma, M. & Samorì, P. Electronic Characterization of Organic Thin Films by Kelvin Probe Force Microscopy. *Adv. Mater.* **18**, 145-164, (2006).

52     Yu, Y.-J. *et al.* Tuning the Graphene Work Function by Electric Field Effect. *Nano Lett.* **9**, 3430-3434, (2009).

53     Matković, A. *et al.* Interfacial Band Engineering of MoS2/Gold Interfaces Using Pyrimidine-Containing Self-Assembled Monolayers: Toward Contact-Resistance-Free Bottom-Contacts. *Advanced Electronic Materials* **6**, (2020).

54     Shlimak, I. *Is Hopping a Science?: Selected Topics of Hopping Conductivity*. (World Scientific Publishing, 2015).

55     Piatti, E. *et al.* Charge transport mechanisms in inkjet-printed thin-film transistors based on two-dimensional materials. *Nature Electronics* **4**, 893-905, (2021).

56     Xu, Q., Yang, G. M. & Zheng, W. T. DFT calculation for stability and quantum capacitance of MoS2 monolayer-based electrode materials. *Materials Today Communications* **22**, (2020).

57     Huo, N. *et al.* High carrier mobility in monolayer CVD-grown MoS2 through phonon suppression. *Nanoscale* **10**, 15071-15077, (2018).

58     Lin, M.-W. *et al.* Thickness-dependent charge transport in few-layer MoS2 field-effect transistors. *Nanotechnology* **27**, (2016).

59     Radisavljevic, B. & Kis, A. Mobility engineering and a metal-insulator transition in monolayer MoS2. *Nat. Mater.* **12**, 815-820, (2013).

60  Sze, S. M. *Semiconductor Devices: Physics and Technology*.  (John Wiley and Sons, 1985).
61  Jariwala, D. *et al.* Band-like transport in high mobility unencapsulated single-layer MoS2 transistors. *Appl. Phys. Lett.* **102**, (2013).
62  Baugher, B. W. H., Churchill, H. O. H., Yang, Y. & Jarillo-Herrero, P. Intrinsic Electronic Transport Properties of High-Quality Monolayer and Bilayer MoS2. *Nano Lett.* **13**, 4212-4216, (2013).
63  Guyot-Sionnest, P. Electrical Transport in Colloidal Quantum Dot Films. *J. Phys. Chem. Lett.* **3**, 1169-1175, (2012).
64  Kim, J. S. *et al.* Electrical Transport Properties of Polymorphic MoS2. *ACS Nano* **10**, 7500-7506, (2016).
65  Qiu, H. *et al.* Hopping transport through defect-induced localized states in molybdenum disulphide. *Nat. Commun.* **4**, (2013).
66  Xue, J. H., Huang, S. Y., Wang, J. Y. & Xu, H. Q. Mott variable-range hopping transport in a MoS2 nanoflake. *RSC Adv.* **9**, 17885-17890, (2019).